\documentclass[sigconf,authorversion,nonacm]{acmart}

\usepackage{amsfonts}
\usepackage{algorithmic}
\usepackage{graphicx}
\usepackage{textcomp}
\usepackage{xcolor}
\usepackage{verbatim}
\usepackage{listings}

\DeclareMathOperator*{\argmin}{ArgMin}

\AtBeginDocument{%
  \providecommand\BibTeX{{%
    \normalfont B\kern-0.5em{\scshape i\kern-0.25em b}\kern-0.8em\TeX}}}

\begin{document}

\title{Dataset: Copy-based Reuse in Open Source Software}

\author{Mahmoud Jahanshahi}
\email{mjahansh@vols.utk.edu}
\affiliation{
  \institution{University of Tennessee}
  \city{Knoxville}
  \country{USA}
}

\author{Audris Mockus}
\email{audris@utk.edu}
\affiliation{
  \institution{University of Tennessee}
  \city{Knoxville}
  \country{USA}
}

\begin{abstract}
In Open Source Software, the source code and any other resources available 
in a project can be viewed or reused by anyone subject to often
permissive licensing restrictions. In
contrast to some studies of dependency-based reuse supported via package managers,
no studies of OSS-wide copy-based reuse exist. This dataset seeks to
encourage the studies of OSS-wide copy-based reuse by providing
copying activity data that captures whole-file copying that captures
nearly all OSS.
To accomplish that, we develop approaches to detect copy-based
reuse by developing an efficient algorithm that exploits World of Code infrastructure: a curated and
cross referenced collection of nearly all open source repositories.
We expect this data will enable future research and tool development
that support such reuse and minimize associated risks.

\end{abstract}


\maketitle

\section{Introduction}
The cornerstone of Open Source Software (OSS) is its ``openness'',
including the ability to access, examine, and copy any project
artifact subject to licensing restrictions that, in turn, often
enforce openness of the derived work. Such ability has a potential to bring dramatic
improvements in developer productivity, but it may also result in
the proliferation of, potentially, poor quality code in various
states of disrepair (e.g, orphan vulnerabilities~\cite{9794064}).

Furthermore, as such copying progresses from one project to another, potentially 
with modifications, critical information about the original design, authorship, 
copyright, and licensing may be lost~\cite{qiu2021empirical}, thus impeding further improvements, 
bug fixing, reducing attribution-related motivation for original authors 
and creating legal problems for downstream users, not only of the copied code 
but of software that depends on at least one package involving copied code~\cite{8667977}. 

As OSS grows, the tasks of tracking the origins of
source code, finding good quality code suitable for reuse, or
untangling parallel evolution of code in multiple projects are ever
more daunting. Despite the longstanding interest, potentially
massive benefits and risks of copying activity, the precise extent,
the prevailing practices, and the potentially negative impacts of
the source code copying at the scale of the entire OSS have not been
investigated mainly due to the challenge of being able to track code over
entire OSS. 

A better understanding of code copying practices may
suggest future research on approaches or tools that make
productivity improvements even greater while, at the same time,
helping to minimize inherent risks of copying.
Specifically, we aim to provide a copy-based reuse dataset to enable 
further analysis of aspects concerning the extent and the nature of  
reuse in OSS and to provide information necessary to investigate approaches that support 
this common activity, make it more efficient, and safer.

First, we create a measurement framework that tracks all versions of source 
code (we refer to a single version as a {\tt blob} in keeping with the 
terminology of the version control system git) across all 
repositories. The time when each unique {\tt blob} $b$ was first
committed to each project $P$ is denoted as $t_b(P)$. The first
repository $P_o(b) = \argmin_Pt_b(P)$ is referred to as the the originating
repository for $b$ (and the first author as the creator). 
Next, copy instances are identified via projects pairs: a project with
the originating commit and the destination project with one of the subsequent commits producing 
the same blob $(P_o(b),P_d(b))$.

\section{Background}
The term software reuse refers to the process of creating software systems 
from existing software rather than building them from 
scratch~\cite{krueger1992software}. Coding from scratch may require more 
time and effort than reusing already created source code that is suitable 
for the task and is of good quality. Programmers, therefore, opportunistically 
reuse code and do so frequently~\cite{juergens2009code}. Programming of 
well-defined problems often start by searching in code repositories followed 
by judicious copying and pasting~\cite{sim1998archetypal}.

Open source software (OSS) development and platforms like GitHub
greatly expand opportunities to reuse by enabling the community of developers to 
curate software projects and by encouraging and enhancing the process of 
opportunistic finding and reusing artifacts.
Much of OSS is specifically designed to be reused and to provide 
resources or functionality to other software 
projects~\cite{haefliger2008code}, thus such reuse can be categorized 
as one of building blocks of OSS. In fact, developers not only look 
for the opportunity to reuse, but also advertise their own 
high quality artifacts for others to reuse~\cite{gharehyazie2017some}. 
High reuse not only brings the software project and its maintainers 
popularity and job prospects~\cite{Roberts06}, but also may bring 
maintainers and corporate support.

\subsection{Research that could be done using this dataset}
Not always is code reuse beneficial though. Although it may reduce the development 
costs, it could pose some other risks that eventually would lead to increased 
overall costs. Security vulnerabilities, licensing and copyright
issues and code quality are just a few such risks~\cite{german2009code}. 
Our curated dataset can enable future research in many potential areas
including but not limited to the following:

\paragraph{Security}
The relation between security and reuse can go both ways: a system can 
become more secure by relying on mature dependencies, or more insecure 
by exposing a larger attack surface via exploitable 
dependencies~\cite{gkortzis2021software}. Specifically, in copy-based 
reuse, extensive code copying in OSS results in an extensive spread of 
potentially vulnerable code not only in inactive projects that are still publicly 
available for others to use that do spread the vulnerability further, 
but also in currently active and in highly popular 
projects~\cite{9794064}. 

\paragraph{Licensing}
As software systems evolve, so do licenses. Various factors, such as 
changes in the legal landscape, commercial code licensed as free and 
open source, or code reused from other open source systems, lead to 
evolution of licensing, which may affect the way a system or part thereof 
can be subsequently used and therefore, it is crucial to monitor licensing
evolution~\cite{di2010exploratory}. But monitoring the huge amount of 
data in entire OSS is not an easy task to do and thus many developers 
fail to adhere to licensing requirements~\cite{an2017stack,german2009license}.

\paragraph{Quality}
Code reuse is not only assumed to inflate maintenance costs in certain 
circumstances, but also considered defect-prone as inconsistent changes 
to code duplicates can lead to unexpected behavior~\cite{juergens2009code}. 
Also forgetting to modify identifiers (variables, functions, 
types, etc.) consistently throughout the reused code will cause errors 
that often slip through compile-time 
checks and become hidden bugs that are very hard to detect~\cite{li2006cp}. 

Other than the bugs introduced by this kind of reuse, the source code 
itself could have bugs or be of low quality that can be spread in the 
same way explained about security vulnerabilities earlier. The
future study using this dataset might suggest approaches
to leverage the information obtained from multiple projects
containing reused code to reduce these kind of risks.

While the described benefits and risks associated with reuse
appear to be real, the extent and the types of copying in the entire
OSS are not clear. In order to prioritize these risks/benefits and
investigate approaches to minimize/maximize them, we
first need to develop an approach to track copy-based reuse scalable
to the massive size of the entire OSS as investigations of convenience
samples presented in prior work do not capture the bulk of copying
activity.

\subsection{Contribution}
To the best of our knowledge no
curation system exists at the level of a {\tt blob}, nor is there an
easy way for anyone to determine the extent of copy-based reuse at
that level and the introduced reuse identification methods (such as
\cite{6975664}) find reuse between given input projects and are not
easily scalable to find reuse across all OSS repositories.  The
methods we use to identify reuse could, therefore,
provide a basis for tools that expose these hard-to-obtain yet
potentially important phenomenon.

Our dataset has two important aspects.
{\it First}, we present the copying activity at the whole open source
software ecosystem level. Previous provided datasets normally focus on
a specific programming language (e.g. Java as in \cite{6624047}) and the
data used in previous works investigating copying
have as well mostly concentrated on a small subset of a specific community (e.g. Java
language, Android apps, etc.) \cite{heinemann2011extent,
haefliger2008code,mockus2007large,hanna2012juxtapp,7958574,sojer2010code} 
or sampled from a single hosting platform (e.g. GitHub)
\cite{gharehyazie2017some, gharehyazie2019cross}. Even research more
comprehensive in programming language
coverage~\cite{lopes2017dejavu} considered only a subset of
programming languages and more importantly, used convenience
sampling by excluding less active
repositories~\cite{10.1109/ICSE43902.2021.00076,9402500}.
Furthermore, almost all research
only focuses on code reuse whereas our dataset tracks all artifacts
whether they are code or other reusable development resources, such
as images or documentation.

{\it Second}, copy-based reuse has not been as extensively investigated as  
the dependency-based reuse, e.g.,~\cite{cox2019surviving,
frakes2001industrial,ossher2010automated}. Copy-based reuse 
is, potentially, no less important, but much less understood form of reuse. In fact,
most of the efforts in copy-based reuse domain are focused on 
clone detection\footnote{identification of, often, relatively small snippets
of code within a single or a limited number of projects} tools and 
techniques~\cite{roy2009comparison,ain2019systematic,
jiang2007deckard,hanna2012juxtapp,white2016deep}, not on the properties
of files that are being reused.
Clone detection tools and techniques usually take a snippet of
code as input and then try to find similar code snippets in a target
directory or an specific domain~\cite{inoue2021finding,svajlenko2013scaling} 
whereas in our dataset, we are finding all
instances of reuse in nearly entirety of OSS.

The description and the curation methods of this dataset has not been published before. 
Furthermore, although the dataset is now publicly available through 
WoC~\footnote{It has been made available only recently}, to the best of our knowledge, 
the data has not been used by authors or others in any published paper yet.

\section{Methodology}\label{s:method}

We start by briefly outlining World of Code infrastructure we employed to 
create our dataset and then present the methods used to identify 
instances of copying.

\subsection{World of Code Infrastructure} \label{WoC} 
Finding duplicate pieces of code and all revisions of that code across all
open source projects is a data and computation intensive task due
to the vast number of OSS projects hosted
on numerous platforms. Previous research on code reuse has, therefore, typically
looked at a relatively small subset of open source software
potentially missing the full extent of copying that could only be
obtained with a nearly complete collection. World of Code
(WoC)~\cite{ma2019world,ma2021world} infrastructure attempts to
remedy this by, on a regular basis, discovering publicly available new and
updated version control repositories, retrieving complete
information (or updates) in them, indexing and cross-referencing
retrieved objects, conducting auto-curation involving author
aliasing~\cite{fry2020dataset} and repository
deforking~\cite{mockus2020complete}, and provides shell, Python and
web APIs to support creation of various research workflows. The
source code version control systems in WoC are collected from
hundreds of forges and, after complete deduplication, takes
approximately 300TB of disk space for the most recent snapshot we
use for our dataset~\footnote{version V}. The specific objective of WoC is to support
research on three kinds of software supply
chains~\cite{ma2018constructing}: technical dependency (traditional
dependency-based package reuse), copy-based reuse, and knowledge
flows~\cite{zhuge2002knowledge,ghobadi2015drives,9463070} (developers
working on, and learning about, projects and then using that
knowledge in their work on other projects).

WoC's operationalization of copy-based supply chains is based on mapping blobs 
(versions of the source code) to all commits and projects where they have been created. 
This implies that copy is detected only if the entire file is copied intact without any 
modifications. Because of that, our dataset includes only the
whole-file copying activity. This also means that different versions of the originally
same file will be considered different objects since they are different blobs.

Furthermore, to discriminate copy-based reuse from forking (a commit uniquely 
identifies modified blobs, and forked projects share commits), we use 
project deforking (p2P map) provided in WoC~\cite{mockus2020complete}. 
Throughout the paper, even if we only use the word project, we mean 
deforked project with the explained definition.

Specifically, WoC uses git object indexing via sha1 signature so that each association has 
to store only the sha1 of the object (in this case blob), and the actual content 
of each object is stored exactly once. When objects are extracted from a repository, 
WoC associates all extracted commits with that repository (the so called c2p map). 
Since a commit points to a tree and to its parent commit objects, the remaining objects 
in a repository can be easily derived by traversing versions and trees. WoC also computes the 
association between commits and blobs created by a commit (new versions of existing files or 
entirely new files) and makes it available via c2fbb map. The map lists all the instances where a 
blob corresponding to one of the files in the repository changed or a new file was created. In the 
former case, the blob corresponding to an earlier version of the file is also provided, 
making it possible to trace back or forth for earlier or newer versions of a blob.

Commits have attributes, such as time of the commit and author of the commit and these 
attributes can be accessed via c2dat map in WoC. A few more maps provided by WoC are also 
used in creating this dataset. 

\begin{figure}[ht]
    \centering
    \includegraphics[width=\linewidth]{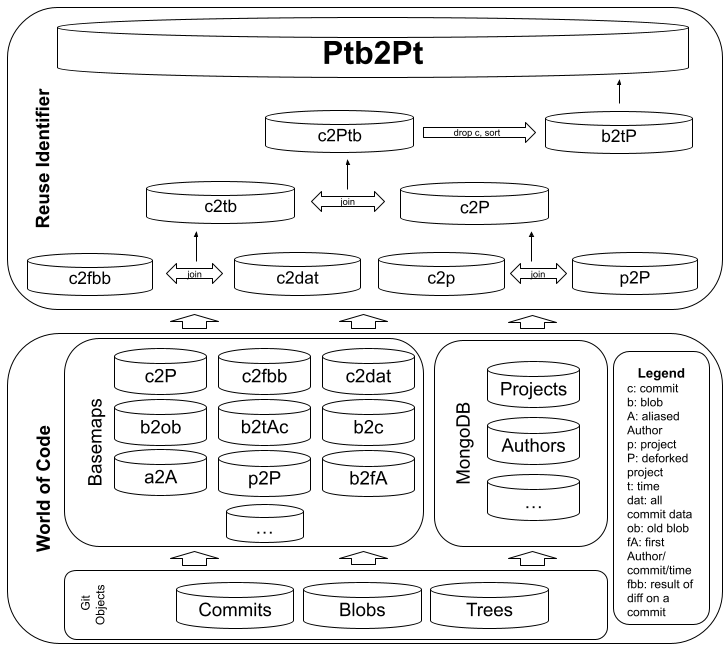}
    \centering
    \caption{Schematic Architecture of Reuse Identification}
    \label{fig:arch}
    \vspace{-.3in}
\end{figure}

\subsection{Identification of reused blobs}\label{s:copied}
Despite the key relationships available in WoC, we have to resolve
several critical obstacles.
We first need to identify the first time $t_b(P)$ each of the nearly 16B blobs
landed in each of the almost 108M projects. We aim to minimize memory
use and be able to run computations in parallel. First, we 
join c2fbb map~\footnote{see https://github.com/woc-hack/tutorial for more information about
WoC map naming convention} (that lists for each commit all the blobs it creates) with
c2dat map (to obtain the date and time of the commit) and then with
the c2P (which itself is the result of joining c2p with p2P maps)
map to identify all projects containing that commit. WoC has
each of the three maps split into 128
partitions\footnote{Partitions are enumerated using the first seven bits of
the sha1 representing the key --- in this case commit --- in order
to obtain partitions of similar size. Each partition is a file sorted by
the key and compressed.} requiring us to
run a sequence of two Unix join commands (first to join c2fbb and
c2dat and then the result of that join with the c2P map) on each of
the 128 partitions in parallel. The result is a new c2Ptb (commit, project,
time, and blob) map stored in 128 partitions $(c^i,P,t,b): i=0,\dots,127$. To create the
timeline for each blob we need to sort all that data by blob,
time, and project. The list has hundreds of billions of rows (20B blobs
often occurring in multiple commits and commits sometimes residing in multiple projects).
We thus needed to break down the problem into smaller pieces to
solve within a reasonable time frame.  
Specifically, we first split each partition $(c^i,P,t,b)$ based on the blob into 128
sub-partitions, thus obtaining 128x128 partitions resulting from the original
partitioning by commits and the secondary one by blobs
$(b^j,t,P,c^i): i,j=0,\dots,127$. 
We then sort each of the 128x128 files by
blob, time, and project (using Unix sort parameterized to handle
extremely large files) and drop all but 
the first commit creating the blob for each project\footnote{A 
blob is often copied within a repository.}.
In the next step we merge 128 commit-based
partitions for each blob-based partition using Unix sort with a merge
option and drop all but the first commit of the blob to a
project. Resulting in 128 blob-based partitions (b2tP map) $(b^j,t,P):
j=0,\dots,127$ where we have only blob, time, and the deforked
project that contain our desired timeline $t_b(P)$. 
Finally, the blob timelines are used to identify instances of copying
$(t_b(P_o), t_b(P_d))$ (or, in the terminology of WoC, Ptb2Pt maps
where the first project is originating\footnote{See section \ref{limit} for the 
limitations in identifying the originating project.} and the second project copied
the blob -- the blob was created at a later time). To accomplish
this we first create a list of blob origination projects and times.
A sweep over b2tP by keeping only the first time and the project associated
with each $b$ and excluding blobs associated with a single
project\footnote{Over 90\% of the blobs belong to a single project, 
so excluding them reduces storage of the relations created downstream.} 
produces $(b^j,t,P_o):j=0,\dots,127$. We also store never reused
blobs $(b^j_{nc},t,P_o):j=0,\dots,127$ (ones that are associated with only one 
project as identified during the sweep mentioned above). 
$(b^j,t,P_o)$ partitions containing only originating project are then
joined with $(b^j,t,P)$ to obtain the cross-product
($(b^j,t_o,P_o,t_d,P_d): j=0,\dots,127, P_o\ne P_d$). Each of the resulting 128 partitions
are then split via project name\footnote{We use the first seven bits of the name's FNV
digest~\cite{fnv} as it is faster and randomizes better short strings than sha1.}, 
into 128 sub-partitions and each sub-partition is then sorted by the originating project:
($(P_o^i,t_o,P_d,t_d,b^j): i,j=0,\dots,127$), then merging over blob-based
partitions belonging to a single project-based partition. Resulting
Ptb2Pt map contains all instances of blob copying: $(t_b(P_o^i),
t_b(P_d))$ and is stored in 128 partitions $i=0,\dots,127$ with each
workflow step described above capable of being run as 128 parallel processes.

\section{Dataset}
The created tables are stored on WoC servers and can be found at
/da?\_data/basemaps/gz/Ptb2PtFullVX.s with X ranging from 0 to 127 
based on the 7 bits in the first byte of the blob sha1. The "V" in
the name indicates that this dataset is based on WoC version 
V\footnote{https://bitbucket.com/swsc/overview}(the latest at the 
time of this work). The format of each file is encoded in its name,
that is, each line of this dataset includes the originating repository
(deforked repository), the timestamp of first commit including the blob
in originating project, blob sha1, destination project (deforked repository)
and the timestamp of first commit including the blob in destination project, 
all separated by semicolon.

\begin{lstlisting}
format:
originating repo;timestamp;blob;
destination repo;timestamp  

example:
MeigeJia_ECE-364;1514098666;
010000001b502dcb0fc8e89d4f854979c93503f8;
HaoboChen1887_Purdue;1598024605
\end{lstlisting}

This means blob 010000001b502dcb0fc8e89d4f854979c93503f8 was first seen in 
MeigeJia\_ECE-364\footnote{Slash symbols are substituted with underscores in WoC
repository naming convention, that is, MeigeJia\_ECE-364 means github.com/MeigeJia/ECE-364. 
Furthermore, the project is hosted on Github unless the domain is mentioned at
the beginning of project name.} repository at 1466402956 (Jun 20 2016) and was 
reused by HaoboChen1887\_Purdue at 1551632725 (Mar 03 2019).

\section{Limitations}\label{limit}

\paragraph{Blob-level reuse}
Our dataset is at entire blob reuse granularity and does not capture the reuse of
pieces of code that form only a part of the file.
Thus blob-level reuse (despite being common) does not represent the
full extent of all code reuse.

Notably, different versions of the same file would have different
blobs as even if two versions differ by only one character, 
they still produce different file hashes (are different blobs).
Thus blob reuse is not the same as file reuse.
File reuse is, however, difficult to define precisely as it is not clear what
files should be considered equivalent in distinct projects.  

\paragraph{Commit time}
The reuse timeline (and identifying the first occurrence) of
a blob is based on the commit timestamp. This time is not always accurate as it depends on
the user's system time. We used suggestions by~\cite{escaping-time-pit} and other 
methods to eliminate incorrect or questionable timestamps. We also used version history 
information to ensure time of parent commits do not postdate that of child commits.

\paragraph{Originating repository}
The accuracy of origination estimates can be increased by the completeness
of data. Even if we assume that the WoC collection is complete, 
some blobs may have been originated in a private repository and then
copied to a public repository, i.e., the originating repository in WoC may 
not be the actual creator of the blob. For example, a 3D cannon pack asset\footnote{
https://assetstore.unity.com/packages/3d/props/weapons/stylish-cannon-pack-174145} was
committed by 38 projects indexed by WoC. That asset, however, was created earlier 
in Unity Asset Store.

\paragraph{Copy instance}
A unique combination of blob, originating project and destination project may 
not always reflect the actual copy pattern because some destination projects 
may have copied the blob not from the originating project (e.g., for projects O, A, and B
in blob creation order, project B may copy either from project O or A).
Also, some blobs are not copied but are created independently in
each repository, e.g, an empty string, or a standard
template automatically created by a common tool. We use the list of such blobs provided
by WoC~\cite{ma2019world} to exclude them from all our calculations.

As was described in each paragraph, we took all the necessary steps to 
minimize the potential negative impact of these limitations and validated the
curated data extensively to ensure its reliability within the boundaries
of limitations.

\section{Future Work}

\paragraph{Dependency-based reuse} 
In this dataset we only introduce the network of copy-based reuse. To better understand reuse in general, it is of great importance to draw a complete picture by creating the reuse network of copy-based
alongside dependency-based reuse as one is not telling the complete 
story without the other.

\paragraph{Code-snippet granularity} 
Another expansion area to better understand reuse is 
setting the granularity to code snippet level. That will produce a much
more complex network of reuse that potentially offers a great
opportunity to have a more in-depth analysis of reuse.

\paragraph{Upstream repository} 
As was mentioned in the limitations section, we still do not exactly know
where a repository copied the blob from and consider it to be the originating repository
in all copy instances. But to better understand how developers find a 
repository to reuse code from, meta heuristic algorithms or artificial intelligence
techniques could be utilized to
predict where the code was copied from in each copy instance.

\clearpage
\bibliographystyle{ACM-Reference-Format}
\bibliography{ref.bib}


\begin{thebibliography}{45}


\ifx \showCODEN    \undefined \def \showCODEN     #1{\unskip}     \fi
\ifx \showDOI      \undefined \def \showDOI       #1{#1}\fi
\ifx \showISBNx    \undefined \def \showISBNx     #1{\unskip}     \fi
\ifx \showISBNxiii \undefined \def \showISBNxiii  #1{\unskip}     \fi
\ifx \showISSN     \undefined \def \showISSN      #1{\unskip}     \fi
\ifx \showLCCN     \undefined \def \showLCCN      #1{\unskip}     \fi
\ifx \shownote     \undefined \def \shownote      #1{#1}          \fi
\ifx \showarticletitle \undefined \def \showarticletitle #1{#1}   \fi
\ifx \showURL      \undefined \def \showURL       {\relax}        \fi
\providecommand\bibfield[2]{#2}
\providecommand\bibinfo[2]{#2}
\providecommand\natexlab[1]{#1}
\providecommand\showeprint[2][]{arXiv:#2}

\bibitem[Ain et~al\mbox{.}(2019)]%
        {ain2019systematic}
\bibfield{author}{\bibinfo{person}{Qurat~Ul Ain}, \bibinfo{person}{Wasi~Haider Butt}, \bibinfo{person}{Muhammad~Waseem Anwar}, \bibinfo{person}{Farooque Azam}, {and} \bibinfo{person}{Bilal Maqbool}.} \bibinfo{year}{2019}\natexlab{}.
\newblock \showarticletitle{A systematic review on code clone detection}.
\newblock \bibinfo{journal}{\emph{IEEE access}}  \bibinfo{volume}{7} (\bibinfo{year}{2019}), \bibinfo{pages}{86121--86144}.
\newblock


\bibitem[An et~al\mbox{.}(2017)]%
        {an2017stack}
\bibfield{author}{\bibinfo{person}{Le An}, \bibinfo{person}{Ons Mlouki}, \bibinfo{person}{Foutse Khomh}, {and} \bibinfo{person}{Giuliano Antoniol}.} \bibinfo{year}{2017}\natexlab{}.
\newblock \showarticletitle{Stack overflow: a code laundering platform?}. In \bibinfo{booktitle}{\emph{2017 IEEE 24th International Conference on Software Analysis, Evolution and Reengineering (SANER)}}. IEEE, \bibinfo{pages}{283--293}.
\newblock


\bibitem[Cox(2019)]%
        {cox2019surviving}
\bibfield{author}{\bibinfo{person}{Russ Cox}.} \bibinfo{year}{2019}\natexlab{}.
\newblock \showarticletitle{Surviving Software Dependencies: Software reuse is finally here but comes with risks.}
\newblock \bibinfo{journal}{\emph{Queue}} \bibinfo{volume}{17}, \bibinfo{number}{2} (\bibinfo{year}{2019}), \bibinfo{pages}{24--47}.
\newblock


\bibitem[Di~Penta et~al\mbox{.}(2010)]%
        {di2010exploratory}
\bibfield{author}{\bibinfo{person}{Massimiliano Di~Penta}, \bibinfo{person}{Daniel~M German}, \bibinfo{person}{Yann-Ga{\"e}l Gu{\'e}h{\'e}neuc}, {and} \bibinfo{person}{Giuliano Antoniol}.} \bibinfo{year}{2010}\natexlab{}.
\newblock \showarticletitle{An exploratory study of the evolution of software licensing}. In \bibinfo{booktitle}{\emph{2010 ACM/IEEE 32nd International Conference on Software Engineering}}, Vol.~\bibinfo{volume}{1}. IEEE, \bibinfo{pages}{145--154}.
\newblock


\bibitem[Feng et~al\mbox{.}(2019)]%
        {8667977}
\bibfield{author}{\bibinfo{person}{Muyue Feng}, \bibinfo{person}{Weixuan Mao}, \bibinfo{person}{Zimu Yuan}, \bibinfo{person}{Yang Xiao}, \bibinfo{person}{Gu Ban}, \bibinfo{person}{Wei Wang}, \bibinfo{person}{Shiyang Wang}, \bibinfo{person}{Qian Tang}, \bibinfo{person}{Jiahuan Xu}, \bibinfo{person}{He Su}, \bibinfo{person}{Binghong Liu}, {and} \bibinfo{person}{Wei Huo}.} \bibinfo{year}{2019}\natexlab{}.
\newblock \showarticletitle{Open-Source License Violations of Binary Software at Large Scale}. In \bibinfo{booktitle}{\emph{2019 IEEE 26th International Conference on Software Analysis, Evolution and Reengineering (SANER)}}. \bibinfo{pages}{564--568}.
\newblock
\urldef\tempurl%
\url{https://doi.org/10.1109/SANER.2019.8667977}
\showDOI{\tempurl}


\bibitem[Fischer et~al\mbox{.}(2017)]%
        {7958574}
\bibfield{author}{\bibinfo{person}{Felix Fischer}, \bibinfo{person}{Konstantin Böttinger}, \bibinfo{person}{Huang Xiao}, \bibinfo{person}{Christian Stransky}, \bibinfo{person}{Yasemin Acar}, \bibinfo{person}{Michael Backes}, {and} \bibinfo{person}{Sascha Fahl}.} \bibinfo{year}{2017}\natexlab{}.
\newblock \showarticletitle{Stack Overflow Considered Harmful? The Impact of Copy\&Paste on Android Application Security}. In \bibinfo{booktitle}{\emph{2017 IEEE Symposium on Security and Privacy (SP)}}. \bibinfo{pages}{121--136}.
\newblock
\urldef\tempurl%
\url{https://doi.org/10.1109/SP.2017.31}
\showDOI{\tempurl}


\bibitem[Flint et~al\mbox{.}(2021)]%
        {escaping-time-pit}
\bibfield{author}{\bibinfo{person}{Samuel~W. Flint}, \bibinfo{person}{Jigyasa Chauhan}, {and} \bibinfo{person}{Robert Dyer}.} \bibinfo{year}{2021}\natexlab{}.
\newblock \showarticletitle{Escaping the Time Pit: Pitfalls and Guidelines for Using Time-Based Git Data}. In \bibinfo{booktitle}{\emph{2021 IEEE/ACM 18th International Conference on Mining Software Repositories (MSR)}}.
\newblock


\bibitem[Frakes and Succi(2001)]%
        {frakes2001industrial}
\bibfield{author}{\bibinfo{person}{William~B Frakes} {and} \bibinfo{person}{Giancarlo Succi}.} \bibinfo{year}{2001}\natexlab{}.
\newblock \showarticletitle{An industrial study of reuse, quality, and productivity}.
\newblock \bibinfo{journal}{\emph{Journal of Systems and Software}} \bibinfo{volume}{57}, \bibinfo{number}{2} (\bibinfo{year}{2001}), \bibinfo{pages}{99--106}.
\newblock


\bibitem[Fry et~al\mbox{.}(2020)]%
        {fry2020dataset}
\bibfield{author}{\bibinfo{person}{Tanner Fry}, \bibinfo{person}{Tapajit Dey}, \bibinfo{person}{Andrey Karnauch}, {and} \bibinfo{person}{Audris Mockus}.} \bibinfo{year}{2020}\natexlab{}.
\newblock \showarticletitle{A dataset and an approach for identity resolution of 38 million author ids extracted from 2b git commits}. In \bibinfo{booktitle}{\emph{Proceedings of the 17th international conference on mining software repositories}}. \bibinfo{pages}{518--522}.
\newblock


\bibitem[German et~al\mbox{.}(2009)]%
        {german2009code}
\bibfield{author}{\bibinfo{person}{Daniel~M German}, \bibinfo{person}{Massimiliano Di~Penta}, \bibinfo{person}{Yann-Gael Gueheneuc}, {and} \bibinfo{person}{Giuliano Antoniol}.} \bibinfo{year}{2009}\natexlab{}.
\newblock \showarticletitle{Code siblings: Technical and legal implications of copying code between applications}. In \bibinfo{booktitle}{\emph{2009 6th IEEE International Working Conference on Mining Software Repositories}}. IEEE, \bibinfo{pages}{81--90}.
\newblock


\bibitem[German and Hassan(2009)]%
        {german2009license}
\bibfield{author}{\bibinfo{person}{Daniel~M German} {and} \bibinfo{person}{Ahmed~E Hassan}.} \bibinfo{year}{2009}\natexlab{}.
\newblock \showarticletitle{License integration patterns: Addressing license mismatches in component-based development}. In \bibinfo{booktitle}{\emph{2009 IEEE 31st international conference on software engineering}}. IEEE, \bibinfo{pages}{188--198}.
\newblock


\bibitem[Gharehyazie et~al\mbox{.}(2017)]%
        {gharehyazie2017some}
\bibfield{author}{\bibinfo{person}{Mohammad Gharehyazie}, \bibinfo{person}{Baishakhi Ray}, {and} \bibinfo{person}{Vladimir Filkov}.} \bibinfo{year}{2017}\natexlab{}.
\newblock \showarticletitle{Some from here, some from there: Cross-project code reuse in github}. In \bibinfo{booktitle}{\emph{2017 IEEE/ACM 14th International Conference on Mining Software Repositories (MSR)}}. IEEE, \bibinfo{pages}{291--301}.
\newblock


\bibitem[Gharehyazie et~al\mbox{.}(2019)]%
        {gharehyazie2019cross}
\bibfield{author}{\bibinfo{person}{Mohammad Gharehyazie}, \bibinfo{person}{Baishakhi Ray}, \bibinfo{person}{Mehdi Keshani}, \bibinfo{person}{Masoumeh~Soleimani Zavosht}, \bibinfo{person}{Abbas Heydarnoori}, {and} \bibinfo{person}{Vladimir Filkov}.} \bibinfo{year}{2019}\natexlab{}.
\newblock \showarticletitle{Cross-project code clones in GitHub}.
\newblock \bibinfo{journal}{\emph{Empirical Software Engineering}} \bibinfo{volume}{24}, \bibinfo{number}{3} (\bibinfo{year}{2019}), \bibinfo{pages}{1538--1573}.
\newblock


\bibitem[Ghobadi(2015)]%
        {ghobadi2015drives}
\bibfield{author}{\bibinfo{person}{Shahla Ghobadi}.} \bibinfo{year}{2015}\natexlab{}.
\newblock \showarticletitle{What drives knowledge sharing in software development teams: A literature review and classification framework}.
\newblock \bibinfo{journal}{\emph{Information \& Management}} \bibinfo{volume}{52}, \bibinfo{number}{1} (\bibinfo{year}{2015}), \bibinfo{pages}{82--97}.
\newblock


\bibitem[Gkortzis et~al\mbox{.}(2021)]%
        {gkortzis2021software}
\bibfield{author}{\bibinfo{person}{Antonios Gkortzis}, \bibinfo{person}{Daniel Feitosa}, {and} \bibinfo{person}{Diomidis Spinellis}.} \bibinfo{year}{2021}\natexlab{}.
\newblock \showarticletitle{Software reuse cuts both ways: An empirical analysis of its relationship with security vulnerabilities}.
\newblock \bibinfo{journal}{\emph{Journal of Systems and Software}}  \bibinfo{volume}{172} (\bibinfo{year}{2021}), \bibinfo{pages}{110653}.
\newblock


\bibitem[Haefliger et~al\mbox{.}(2008)]%
        {haefliger2008code}
\bibfield{author}{\bibinfo{person}{Stefan Haefliger}, \bibinfo{person}{Georg Von~Krogh}, {and} \bibinfo{person}{Sebastian Spaeth}.} \bibinfo{year}{2008}\natexlab{}.
\newblock \showarticletitle{Code reuse in open source software}.
\newblock \bibinfo{journal}{\emph{Management science}} \bibinfo{volume}{54}, \bibinfo{number}{1} (\bibinfo{year}{2008}), \bibinfo{pages}{180--193}.
\newblock


\bibitem[Hanna et~al\mbox{.}(2012)]%
        {hanna2012juxtapp}
\bibfield{author}{\bibinfo{person}{Steve Hanna}, \bibinfo{person}{Ling Huang}, \bibinfo{person}{Edward Wu}, \bibinfo{person}{Saung Li}, \bibinfo{person}{Charles Chen}, {and} \bibinfo{person}{Dawn Song}.} \bibinfo{year}{2012}\natexlab{}.
\newblock \showarticletitle{Juxtapp: A scalable system for detecting code reuse among android applications}. In \bibinfo{booktitle}{\emph{International Conference on Detection of Intrusions and Malware, and Vulnerability Assessment}}. Springer, \bibinfo{pages}{62--81}.
\newblock


\bibitem[Hata et~al\mbox{.}(2021a)]%
        {9402500}
\bibfield{author}{\bibinfo{person}{Hideaki Hata}, \bibinfo{person}{Raula Gaikovina~Kula}, \bibinfo{person}{Takashi Ishio}, {and} \bibinfo{person}{Christoph Treude}.} \bibinfo{year}{2021}\natexlab{a}.
\newblock \showarticletitle{Research Artifact: The Potential of Meta-Maintenance on GitHub}. In \bibinfo{booktitle}{\emph{2021 IEEE/ACM 43rd International Conference on Software Engineering: Companion Proceedings (ICSE-Companion)}}. \bibinfo{pages}{192--193}.
\newblock
\urldef\tempurl%
\url{https://doi.org/10.1109/ICSE-Companion52605.2021.00084}
\showDOI{\tempurl}


\bibitem[Hata et~al\mbox{.}(2021b)]%
        {10.1109/ICSE43902.2021.00076}
\bibfield{author}{\bibinfo{person}{Hideaki Hata}, \bibinfo{person}{Raula~Gaikovina Kula}, \bibinfo{person}{Takashi Ishio}, {and} \bibinfo{person}{Christoph Treude}.} \bibinfo{year}{2021}\natexlab{b}.
\newblock \showarticletitle{Same File, Different Changes: The Potential of Meta-Maintenance on GitHub}. In \bibinfo{booktitle}{\emph{Proceedings of the 43rd International Conference on Software Engineering}} (Madrid, Spain) \emph{(\bibinfo{series}{ICSE '21})}. \bibinfo{publisher}{IEEE Press}, \bibinfo{pages}{773–784}.
\newblock
\showISBNx{9781450390859}
\urldef\tempurl%
\url{https://doi.org/10.1109/ICSE43902.2021.00076}
\showDOI{\tempurl}


\bibitem[Heinemann et~al\mbox{.}(2011)]%
        {heinemann2011extent}
\bibfield{author}{\bibinfo{person}{Lars Heinemann}, \bibinfo{person}{Florian Deissenboeck}, \bibinfo{person}{Mario Gleirscher}, \bibinfo{person}{Benjamin Hummel}, {and} \bibinfo{person}{Maximilian Irlbeck}.} \bibinfo{year}{2011}\natexlab{}.
\newblock \showarticletitle{On the extent and nature of software reuse in open source java projects}. In \bibinfo{booktitle}{\emph{International Conference on Software Reuse}}. Springer, \bibinfo{pages}{207--222}.
\newblock


\bibitem[Inoue et~al\mbox{.}(2021)]%
        {inoue2021finding}
\bibfield{author}{\bibinfo{person}{Katsuro Inoue}, \bibinfo{person}{Yuya Miyamoto}, \bibinfo{person}{Daniel~M German}, {and} \bibinfo{person}{Takashi Ishio}.} \bibinfo{year}{2021}\natexlab{}.
\newblock \showarticletitle{Finding code-clone snippets in large source-code collection by CCgrep}. In \bibinfo{booktitle}{\emph{Open Source Systems: 17th IFIP WG 2.13 International Conference, OSS 2021, Virtual Event, May 12--13, 2021, Proceedings 17}}. Springer, \bibinfo{pages}{28--41}.
\newblock


\bibitem[Janjic et~al\mbox{.}(2013)]%
        {6624047}
\bibfield{author}{\bibinfo{person}{Werner Janjic}, \bibinfo{person}{Oliver Hummel}, \bibinfo{person}{Marcus Schumacher}, {and} \bibinfo{person}{Colin Atkinson}.} \bibinfo{year}{2013}\natexlab{}.
\newblock \showarticletitle{An unabridged source code dataset for research in software reuse}. In \bibinfo{booktitle}{\emph{2013 10th Working Conference on Mining Software Repositories (MSR)}}. \bibinfo{pages}{339--342}.
\newblock
\urldef\tempurl%
\url{https://doi.org/10.1109/MSR.2013.6624047}
\showDOI{\tempurl}


\bibitem[Jiang et~al\mbox{.}(2007)]%
        {jiang2007deckard}
\bibfield{author}{\bibinfo{person}{Lingxiao Jiang}, \bibinfo{person}{Ghassan Misherghi}, \bibinfo{person}{Zhendong Su}, {and} \bibinfo{person}{Stephane Glondu}.} \bibinfo{year}{2007}\natexlab{}.
\newblock \showarticletitle{Deckard: Scalable and accurate tree-based detection of code clones}. In \bibinfo{booktitle}{\emph{29th International Conference on Software Engineering (ICSE'07)}}. IEEE, \bibinfo{pages}{96--105}.
\newblock


\bibitem[Juergens et~al\mbox{.}(2009)]%
        {juergens2009code}
\bibfield{author}{\bibinfo{person}{Elmar Juergens}, \bibinfo{person}{Florian Deissenboeck}, \bibinfo{person}{Benjamin Hummel}, {and} \bibinfo{person}{Stefan Wagner}.} \bibinfo{year}{2009}\natexlab{}.
\newblock \showarticletitle{Do code clones matter?}. In \bibinfo{booktitle}{\emph{2009 IEEE 31st International Conference on Software Engineering}}. IEEE, \bibinfo{pages}{485--495}.
\newblock


\bibitem[Kawamitsu et~al\mbox{.}(2014)]%
        {6975664}
\bibfield{author}{\bibinfo{person}{Naohiro Kawamitsu}, \bibinfo{person}{Takashi Ishio}, \bibinfo{person}{Tetsuya Kanda}, \bibinfo{person}{Raula~Gaikovina Kula}, \bibinfo{person}{Coen De~Roover}, {and} \bibinfo{person}{Katsuro Inoue}.} \bibinfo{year}{2014}\natexlab{}.
\newblock \showarticletitle{Identifying Source Code Reuse across Repositories Using LCS-Based Source Code Similarity}. In \bibinfo{booktitle}{\emph{2014 IEEE 14th International Working Conference on Source Code Analysis and Manipulation}}. \bibinfo{pages}{305--314}.
\newblock
\urldef\tempurl%
\url{https://doi.org/10.1109/SCAM.2014.17}
\showDOI{\tempurl}


\bibitem[Krueger(1992)]%
        {krueger1992software}
\bibfield{author}{\bibinfo{person}{Charles~W Krueger}.} \bibinfo{year}{1992}\natexlab{}.
\newblock \showarticletitle{Software reuse}.
\newblock \bibinfo{journal}{\emph{ACM Computing Surveys (CSUR)}} \bibinfo{volume}{24}, \bibinfo{number}{2} (\bibinfo{year}{1992}), \bibinfo{pages}{131--183}.
\newblock


\bibitem[Li et~al\mbox{.}(2006)]%
        {li2006cp}
\bibfield{author}{\bibinfo{person}{Zhenmin Li}, \bibinfo{person}{Shan Lu}, \bibinfo{person}{Suvda Myagmar}, {and} \bibinfo{person}{Yuanyuan Zhou}.} \bibinfo{year}{2006}\natexlab{}.
\newblock \showarticletitle{CP-Miner: Finding copy-paste and related bugs in large-scale software code}.
\newblock \bibinfo{journal}{\emph{IEEE Transactions on software Engineering}} \bibinfo{volume}{32}, \bibinfo{number}{3} (\bibinfo{year}{2006}), \bibinfo{pages}{176--192}.
\newblock


\bibitem[Lopes et~al\mbox{.}(2017)]%
        {lopes2017dejavu}
\bibfield{author}{\bibinfo{person}{Cristina~V Lopes}, \bibinfo{person}{Petr Maj}, \bibinfo{person}{Pedro Martins}, \bibinfo{person}{Vaibhav Saini}, \bibinfo{person}{Di Yang}, \bibinfo{person}{Jakub Zitny}, \bibinfo{person}{Hitesh Sajnani}, {and} \bibinfo{person}{Jan Vitek}.} \bibinfo{year}{2017}\natexlab{}.
\newblock \showarticletitle{D{\'e}j{\`a}Vu: a map of code duplicates on GitHub}.
\newblock \bibinfo{journal}{\emph{Proceedings of the ACM on Programming Languages}} \bibinfo{volume}{1}, \bibinfo{number}{OOPSLA} (\bibinfo{year}{2017}), \bibinfo{pages}{1--28}.
\newblock


\bibitem[Lyulina and Jahanshahi(2021)]%
        {9463070}
\bibfield{author}{\bibinfo{person}{Elena Lyulina} {and} \bibinfo{person}{Mahmoud Jahanshahi}.} \bibinfo{year}{2021}\natexlab{}.
\newblock \showarticletitle{Building the Collaboration Graph of Open-Source Software Ecosystem}. In \bibinfo{booktitle}{\emph{2021 IEEE/ACM 18th International Conference on Mining Software Repositories (MSR)}}. \bibinfo{pages}{618--620}.
\newblock
\urldef\tempurl%
\url{https://doi.org/10.1109/MSR52588.2021.00086}
\showDOI{\tempurl}


\bibitem[Ma(2018)]%
        {ma2018constructing}
\bibfield{author}{\bibinfo{person}{Yuxing Ma}.} \bibinfo{year}{2018}\natexlab{}.
\newblock \showarticletitle{Constructing supply chains in open source software}. In \bibinfo{booktitle}{\emph{2018 IEEE/ACM 40th International Conference on Software Engineering: Companion (ICSE-Companion)}}. IEEE, \bibinfo{pages}{458--459}.
\newblock


\bibitem[Ma et~al\mbox{.}(2019)]%
        {ma2019world}
\bibfield{author}{\bibinfo{person}{Yuxing Ma}, \bibinfo{person}{Chris Bogart}, \bibinfo{person}{Sadika Amreen}, \bibinfo{person}{Russell Zaretzki}, {and} \bibinfo{person}{Audris Mockus}.} \bibinfo{year}{2019}\natexlab{}.
\newblock \showarticletitle{World of code: an infrastructure for mining the universe of open source VCS data}. In \bibinfo{booktitle}{\emph{2019 IEEE/ACM 16th International Conference on Mining Software Repositories (MSR)}}. IEEE, \bibinfo{pages}{143--154}.
\newblock


\bibitem[Ma et~al\mbox{.}(2021)]%
        {ma2021world}
\bibfield{author}{\bibinfo{person}{Yuxing Ma}, \bibinfo{person}{Tapajit Dey}, \bibinfo{person}{Chris Bogart}, \bibinfo{person}{Sadika Amreen}, \bibinfo{person}{Marat Valiev}, \bibinfo{person}{Adam Tutko}, \bibinfo{person}{David Kennard}, \bibinfo{person}{Russell Zaretzki}, {and} \bibinfo{person}{Audris Mockus}.} \bibinfo{year}{2021}\natexlab{}.
\newblock \showarticletitle{World of code: Enabling a research workflow for mining and analyzing the universe of open source vcs data}.
\newblock \bibinfo{journal}{\emph{Empirical Software Engineering}} \bibinfo{volume}{26}, \bibinfo{number}{2} (\bibinfo{year}{2021}), \bibinfo{pages}{1--42}.
\newblock


\bibitem[Mockus(2007)]%
        {mockus2007large}
\bibfield{author}{\bibinfo{person}{Audris Mockus}.} \bibinfo{year}{2007}\natexlab{}.
\newblock \showarticletitle{Large-scale code reuse in open source software}. In \bibinfo{booktitle}{\emph{First International Workshop on Emerging Trends in FLOSS Research and Development (FLOSS'07: ICSE Workshops 2007)}}. IEEE, \bibinfo{pages}{7--7}.
\newblock


\bibitem[Mockus et~al\mbox{.}(2020)]%
        {mockus2020complete}
\bibfield{author}{\bibinfo{person}{Audris Mockus}, \bibinfo{person}{Diomidis Spinellis}, \bibinfo{person}{Zoe Kotti}, {and} \bibinfo{person}{Gabriel~John Dusing}.} \bibinfo{year}{2020}\natexlab{}.
\newblock \showarticletitle{A complete set of related git repositories identified via community detection approaches based on shared commits}. In \bibinfo{booktitle}{\emph{Proceedings of the 17th International Conference on Mining Software Repositories}}. \bibinfo{pages}{513--517}.
\newblock


\bibitem[Noll(2012)]%
        {fnv}
\bibfield{author}{\bibinfo{person}{Landon~Curt Noll}.} \bibinfo{year}{2012}\natexlab{}.
\newblock \showarticletitle{Fowler/noll/vo (fnv) hash}.
\newblock \bibinfo{journal}{\emph{Accessed Jan}} (\bibinfo{year}{2012}).
\newblock


\bibitem[Ossher et~al\mbox{.}(2010)]%
        {ossher2010automated}
\bibfield{author}{\bibinfo{person}{Joel Ossher}, \bibinfo{person}{Sushil Bajracharya}, {and} \bibinfo{person}{Cristina Lopes}.} \bibinfo{year}{2010}\natexlab{}.
\newblock \showarticletitle{Automated dependency resolution for open source software}. In \bibinfo{booktitle}{\emph{2010 7th IEEE Working Conference on Mining Software Repositories (MSR 2010)}}. IEEE, \bibinfo{pages}{130--140}.
\newblock


\bibitem[Qiu et~al\mbox{.}(2021)]%
        {qiu2021empirical}
\bibfield{author}{\bibinfo{person}{Shi Qiu}, \bibinfo{person}{Daniel~M German}, {and} \bibinfo{person}{Katsuro Inoue}.} \bibinfo{year}{2021}\natexlab{}.
\newblock \showarticletitle{Empirical study on dependency-related license violation in the javascript package ecosystem}.
\newblock \bibinfo{journal}{\emph{Journal of Information Processing}}  \bibinfo{volume}{29} (\bibinfo{year}{2021}), \bibinfo{pages}{296--304}.
\newblock


\bibitem[Reid et~al\mbox{.}(2022)]%
        {9794064}
\bibfield{author}{\bibinfo{person}{David Reid}, \bibinfo{person}{Mahmoud Jahanshahi}, {and} \bibinfo{person}{Audris Mockus}.} \bibinfo{year}{2022}\natexlab{}.
\newblock \showarticletitle{The Extent of Orphan Vulnerabilities from Code Reuse in Open Source Software}. In \bibinfo{booktitle}{\emph{2022 IEEE/ACM 44th International Conference on Software Engineering (ICSE)}}. \bibinfo{pages}{2104--2115}.
\newblock
\urldef\tempurl%
\url{https://doi.org/10.1145/3510003.3510216}
\showDOI{\tempurl}


\bibitem[Roberts et~al\mbox{.}(2006)]%
        {Roberts06}
\bibfield{author}{\bibinfo{person}{Jeffrey~A. Roberts}, \bibinfo{person}{Il-Horn Hann}, {and} \bibinfo{person}{Sandra~A. Slaughter}.} \bibinfo{year}{2006}\natexlab{}.
\newblock \showarticletitle{Understanding the motivations, participation, and performance of open source software developers: A longitudinal study of the apache projects}.
\newblock \bibinfo{journal}{\emph{Management Science}} \bibinfo{volume}{52}, \bibinfo{number}{7} (\bibinfo{date}{July} \bibinfo{year}{2006}), \bibinfo{pages}{984--999}.
\newblock


\bibitem[Roy et~al\mbox{.}(2009)]%
        {roy2009comparison}
\bibfield{author}{\bibinfo{person}{Chanchal~K Roy}, \bibinfo{person}{James~R Cordy}, {and} \bibinfo{person}{Rainer Koschke}.} \bibinfo{year}{2009}\natexlab{}.
\newblock \showarticletitle{Comparison and evaluation of code clone detection techniques and tools: A qualitative approach}.
\newblock \bibinfo{journal}{\emph{Science of computer programming}} \bibinfo{volume}{74}, \bibinfo{number}{7} (\bibinfo{year}{2009}), \bibinfo{pages}{470--495}.
\newblock


\bibitem[Sim et~al\mbox{.}(1998)]%
        {sim1998archetypal}
\bibfield{author}{\bibinfo{person}{Susan~Elliott Sim}, \bibinfo{person}{Charles~LA Clarke}, {and} \bibinfo{person}{Richard~C Holt}.} \bibinfo{year}{1998}\natexlab{}.
\newblock \showarticletitle{Archetypal source code searches: A survey of software developers and maintainers}. In \bibinfo{booktitle}{\emph{Proceedings. 6th International Workshop on Program Comprehension. IWPC'98 (Cat. No. 98TB100242)}}. IEEE, \bibinfo{pages}{180--187}.
\newblock


\bibitem[Sojer and Henkel(2010)]%
        {sojer2010code}
\bibfield{author}{\bibinfo{person}{Manuel Sojer} {and} \bibinfo{person}{Joachim Henkel}.} \bibinfo{year}{2010}\natexlab{}.
\newblock \showarticletitle{Code reuse in open source software development: Quantitative evidence, drivers, and impediments}.
\newblock \bibinfo{journal}{\emph{Journal of the Association for Information Systems}} \bibinfo{volume}{11}, \bibinfo{number}{12} (\bibinfo{year}{2010}), \bibinfo{pages}{2}.
\newblock


\bibitem[Svajlenko et~al\mbox{.}(2013)]%
        {svajlenko2013scaling}
\bibfield{author}{\bibinfo{person}{Jeffrey Svajlenko}, \bibinfo{person}{Iman Keivanloo}, {and} \bibinfo{person}{Chanchal~K Roy}.} \bibinfo{year}{2013}\natexlab{}.
\newblock \showarticletitle{Scaling classical clone detection tools for ultra-large datasets: An exploratory study}. In \bibinfo{booktitle}{\emph{2013 7th International Workshop on Software Clones (IWSC)}}. IEEE, \bibinfo{pages}{16--22}.
\newblock


\bibitem[White et~al\mbox{.}(2016)]%
        {white2016deep}
\bibfield{author}{\bibinfo{person}{Martin White}, \bibinfo{person}{Michele Tufano}, \bibinfo{person}{Christopher Vendome}, {and} \bibinfo{person}{Denys Poshyvanyk}.} \bibinfo{year}{2016}\natexlab{}.
\newblock \showarticletitle{Deep learning code fragments for code clone detection}. In \bibinfo{booktitle}{\emph{2016 31st IEEE/ACM International Conference on Automated Software Engineering (ASE)}}. IEEE, \bibinfo{pages}{87--98}.
\newblock


\bibitem[Zhuge(2002)]%
        {zhuge2002knowledge}
\bibfield{author}{\bibinfo{person}{Hai Zhuge}.} \bibinfo{year}{2002}\natexlab{}.
\newblock \showarticletitle{Knowledge flow management for distributed team software development}.
\newblock \bibinfo{journal}{\emph{Knowledge-Based Systems}} \bibinfo{volume}{15}, \bibinfo{number}{8} (\bibinfo{year}{2002}), \bibinfo{pages}{465--471}.
\newblock


\end{thebibliography}

\end{document}